\documentclass[aps,prb,reprint]{revtex4-1}
\usepackage{amsmath}
\usepackage{amssymb}
\usepackage{braket}
\usepackage{graphicx}
\usepackage[tabele]{xcolor}
\usepackage{verbatim}
\usepackage{float}
\usepackage{bbold}
\usepackage{multirow}
\begin{document}
\title{Hybrid $\mathbf{k\cdot p}$-tight-binding model for intersubband optics in atomically thin InSe films}
\author{S. J. Magorrian}
\author{A. Ceferino}
\author{V. Z\'{o}lyomi}
\author{V. I. Fal'ko}
\affiliation{National Graphene Institute, University of Manchester, Booth St E, Manchester, M13 9PL, United Kingdom}
\affiliation{School of Physics and Astronomy, University of Manchester, Oxford Road, Manchester, M13 9PL, United Kingdom}
\begin{abstract}
We propose atomic films of n-doped $\gamma$-InSe as a platform for intersubband optics in the infrared (IR) and far infrared (FIR) range, coupled to out-of-plane polarized light. Depending on the film thickness (number of layers) and amount of n-doping of the InSe film these transitions span from $\sim 0.7$ eV for bilayer to $\sim 0.05$ eV for 15-layer InSe. We use a hybrid $\mathbf{k} \cdot \mathbf{p}$ theory and tight-binding model, fully parametrized using density functional theory, to predict their oscillator strengths and thermal linewidths at room temperature.
\end{abstract}
\maketitle
\section{Introduction}
Atomically thin layers of van der Waals (vdW) materials and their heterostructures\cite{Novoselov2012,Geim2013}, generally branded as two-dimensional materials (2DMs), came to the spotlight due to their promise for creating multifunctional electronic devices and, more generally, as a new materials-base for optoelectronics\cite{Ferrari2015}. This class of materials features strong covalent bonding of atoms in the 2D planes and a weak vdW attraction between the layers, permitting fabrication of stable films of such materials down to monolayer (sub-nm) thickness and creation of their various heterostructures. The ongoing studies of 2DMs broadly address graphene\cite{Novoselov2012} and hexagonal boron nitride (hBN, a wide band gap insulator)\cite{Gorbachev2011}, narrow-gap semiconductor black phosphorus\cite{Li2014,Liu2014}, and various transition metal dichalcogenides\cite{Wang2012}. 

Among all 2DMs, a special place is taken by two post-transition metal chalcogenides (PTMCs):  InSe and GaSe. This closely lattice-matched pair of optically active 2D compounds (with a monolayer stoichiometric formula M$_2$Se$_2$, M$=$In or Ga) was found, both theoretically\cite{Zolyomi2014,magorrian2016electronic,*tb_erratum} and experimentally\cite{bandurin2017high}, to have a band gap that varies widely from the monolayer to multilayer films, densely covering the range of energies $E_g\sim1.3  - 3$~eV.  Also, these 2DMs have relatively light ($m_c\sim 0.2m_e$) conduction band electrons\cite{Zolyomi2014,magorrian2016electronic,bandurin2017high} with very high mobility, even in the case of atomically thin films. While the recent optical studies of 2D InSe and GaSe crystals\cite{bandurin2017high,GaSe_PL} have been performed using mechanically exfoliated films, manufacturability of 2D crystals of PTMCs using molecular beam epitaxy\cite{BenAziza} and chemical vapour deposition\cite{Browning2017} has already been demonstrated, and the potential of various PTMCs for optoelectronics applications identified in terms of their implementation in high-sensitivity\cite{Tamalampudi2014} and fast\cite{Balakrishnan2014} broad-band photodiodes. Here, we show that optical transitions between subbands in n-doped PTMC films of various thicknesses, active in the same out-of-plane polarization\cite{magorrian2016electronic} as the interband transitions, can extend the range of their optical functionality into the IR/FIR range.

Theoretical studies of 2D InSe have largely focused on the monolayer, with DFT studies finding a slightly indirect band gap due to an offset in the valence band maximum\cite{Zolyomi2014}, with a Lifshitz transition presenting the possibility of ferromagnetism on hole-doping\cite{Louie2015}. Meanwhile, $\mathbf{k\cdot p}$ theory and tight-binding studies\cite{Appelbaum2015,magorrian2016electronic,zhou2017multiband} have been employed to further understand the band structure, symmetries, optical properties, and highly sensitive strain response of monolayer InSe.
\begin{figure}
	\centering
	\includegraphics[width=0.95\linewidth]{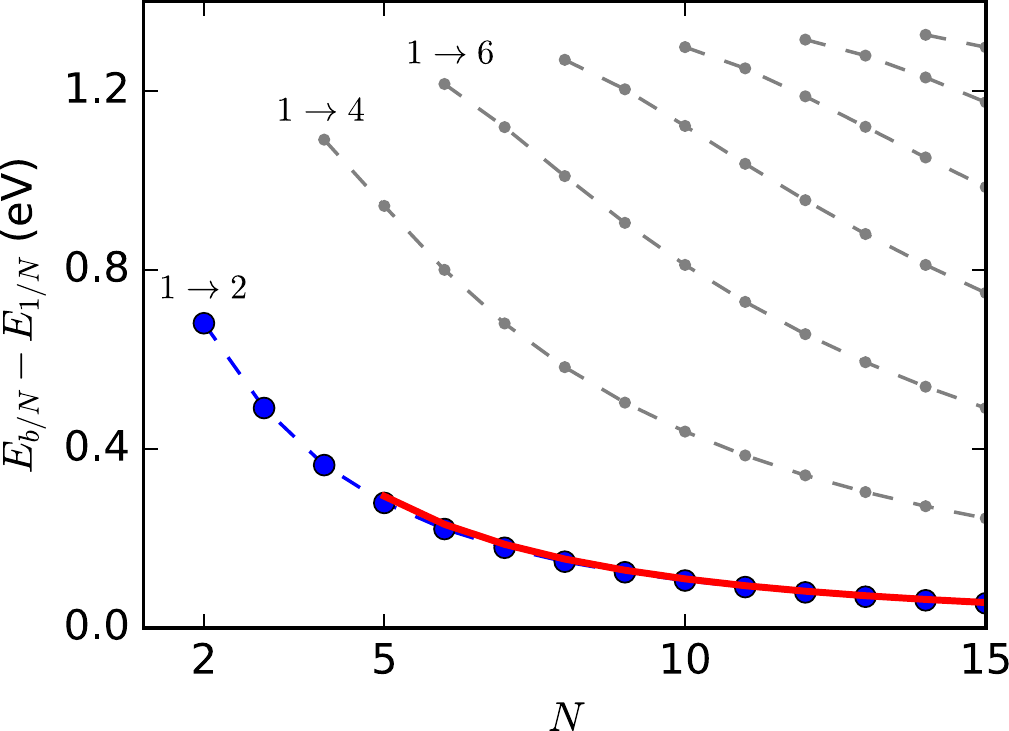}
	\caption{Intersubband energies for allowed electric dipole transition for excitation from the lowest sub-band in weakly n-doped $N=2-15$-layer InSe. Transitions to the second lowest sub-band (marked in blue) are expected to be significantly stronger than transitions to higher sub-bands. The red line shows the $1|N\rightarrow 2|N$ intersubband transition energies in lightly n-doped films approximated by an asymptotic ($N\gg 1$) formula, $\hbar\omega\approx\frac{\hbar^2\pi^2}{2m_{A z}a_z^2}\frac{3}{(N+2\nu)^2}$, derived from Eq. (\ref{eq:me_model_12}). The lowest intersubband transition energy increases for heavily doped films (see Fig. \ref{fig:screen} in Sec. V).}
	\label{fig:lw}
\end{figure}

Here, we use a hybrid $\mathbf{k\cdot p}$ theory and tight-binding (HkpTB) model to study in detail the subbands and intersubband transitions in atomically thin films of post-transition metal chalcogenides. In particular we find that, in InSe films with thicknesses from $N=2$ to $N=15$ layers, transitions between the lowest and first excited subbands cover the range of photons from $\lambda\sim 2~\mu\mathrm{m}$ to $\lambda\sim 25~\mu\mathrm{m}$ (between $\sim$680~meV and $\sim$50~meV), Fig. \ref{fig:lw}. We analyze thermal broadening of the intersubband absorption spectra caused by the variation of the 2D (in-plane) dispersion of electrons in consecutive subbands, and we also develop the self-consistent description of the subband energies for the films doped n-type by gates.

\section{Hybrid $\mathbf{k\cdot p}$-tight-binding model}

\begin{figure}
	\centering
	\includegraphics[width=0.95\linewidth]{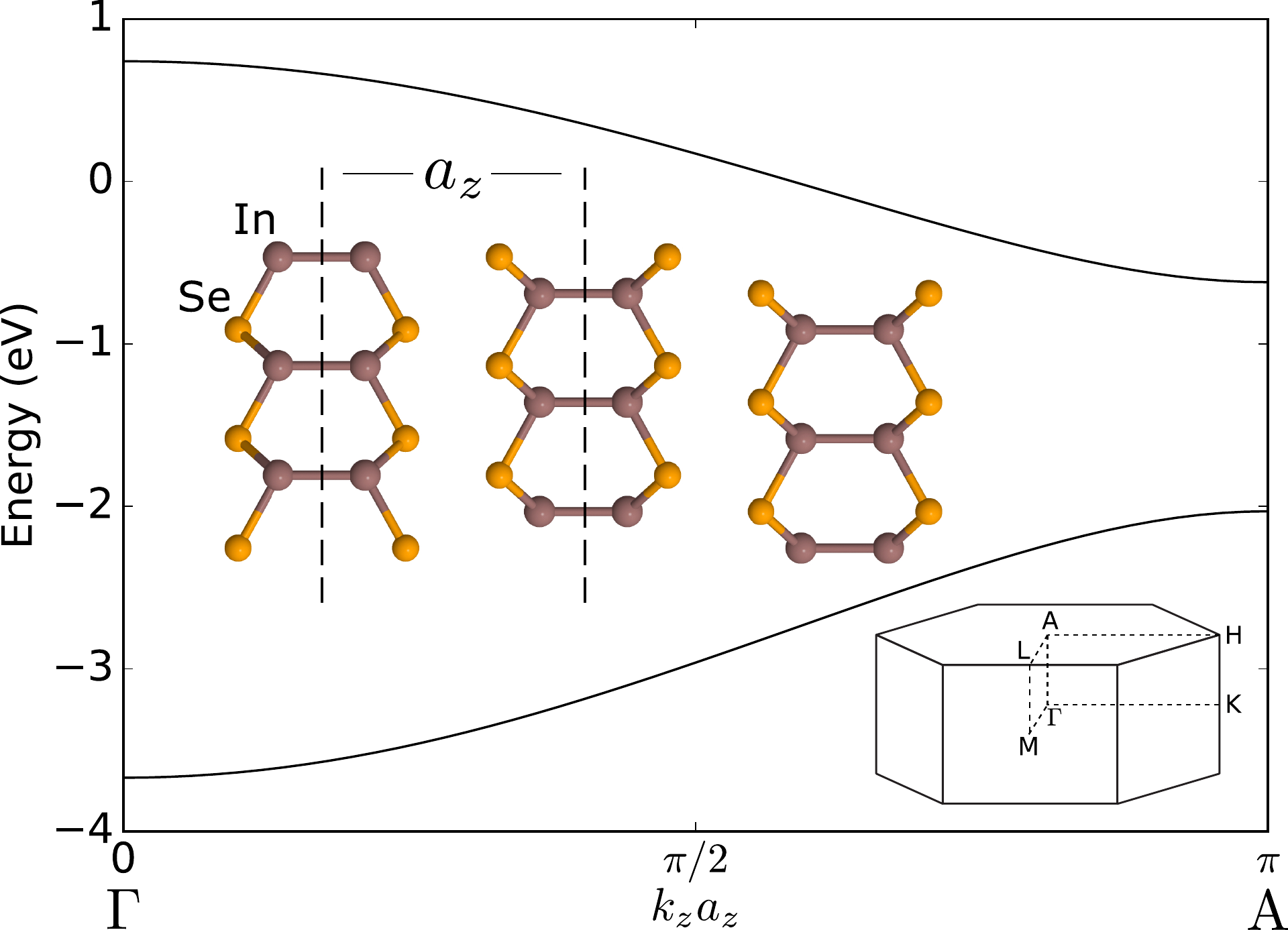}
	\caption{$\Gamma$-A dispersion in bulk $\gamma$-InSe ($k_x=k_y=0$), from two-band HkpTB model, Eq. (\ref{full_ham}). Zero of energy set to conduction band edge in monolayer. Inset center - crystal structure of $\gamma$-InSe. Monolayer has a hexagonal structure, with point group symmetry $D_{3h}$. The point group of the bulk crystal is $C_{3v}$, with each layer shifted with respect to the layer below such that selenium atoms in the upper layer lie above the indium atoms in the lower layer. $a_z=8.32$~\AA~is the experimentally known interlayer distance\cite{Mudd2013}. Inset bottom right - Brillouin zone of conventional unit cell of InSe (3 layers), bands plotted here have been unfolded.}
	\label{fig:crys}
\end{figure}
The crystal structure of few-layer InSe is shown in Fig. \ref{fig:crys}, with successive Se-In-In-Se layers arranged in the $\gamma$ polytype -- each layer is shifted with respect to the layer below such that selenium atoms in the upper layer lie above the indium atoms in the lower layer. The wavefunctions at the conduction band edge in InSe are predominantly composed of $s$ and $p_z$ orbitals on In and Se\cite{magorrian2016electronic,*tb_erratum}. Electrons in the monolayer have a light in-plane effective mass $m_c\sim 0.2~m_e$, while strong interlayer hopping between the layers leads to a strong band gap dependence on the number of layers, varying from $\sim 1.3$~eV in the bulk to $\sim 2.0$~eV in the bilayer\cite{bandurin2017high,magorrian2016electronic}.

To describe subbands of electrons in the conduction band in few-layer InSe we construct a 2-band hybrid $\mathbf{k\cdot p}$-tight-binding Hamiltonian in a basis of the $\mathbf{k\cdot p}$ conduction and valence bands of the monolayer, with successive layers coupled by tight-binding hoppings between monolayer $\mathbf{k\cdot p}$ states. These bands and hoppings are chosen as those in the region of the band edge with non-negligible strength interlayer electronic couplings and subband splittings. The Hamiltonian takes the form

	\begin{align}
	\label{full_ham}
	\hat{H}=\sum_{n}^N&\bigg[\left(\Delta_c(2-\delta_{n,1}-\delta_{n,N})+\frac{\hbar^2 p^2}{2m_c}\right)a_{nc}a_{nc}^{\dagger}\\
	&+(E_v+\Delta_v(2-\delta_{n,1}-\delta_{n,N}))a_{nv}a_{nv}^{\dagger}\bigg]\nonumber\\
	+\sum_{n}^{N-1}&\left[(t_c^{\Gamma}+t_c^{\prime}p^2)a^{\dagger}_{(n+1)c}a_{nc}+t_{v}a^{\dagger}_{(n+1)v}a_{nv}\right.\nonumber\\ 
	&\left.+(t_{cv}^{\Gamma}+t_{cv}^{\prime}p^2)\left(a^{\dagger}_{(n+1)v}a_{nc}-a^{\dagger}_{(n+1)c}a_{nv}\right)+\mathrm{H.c.}\right]\nonumber.
	\end{align}
Here, operators $a^{(\dagger)}_{nc/v}$ annihilate (create) electrons in the conduction/valence bands of the individual layers (indexed by $n=1,...,N$) of the $N$-layer crystal. Since the $\Gamma$-point dispersion of electrons in  the conduction band of monolayer InSe changes negligibly on inclusion of spin-orbit coupling (SOC)\cite{soc2017} we neglect spin-orbit effects and describe the monolayer conduction band with a parabolic dispersion with effective mass $m_c$, while approximating the valence band as flat, with constant energy $E_v$. $t_{c(v)}$ is an interlayer conduction-conduction (valence-valence) hop, and $t_{cv}$ describes interlayer conduction-valence and valence-conduction hybridization. Our earlier studies\cite{magorrian2016electronic} showed that the interlayer coupling is dominated by Se-Se interlayer pairs on the outside adjacent monolayers, and hence we assume that the valence-conduction and conduction-valence hops can be related as $t_{vc}=-t_{cv}$. The $p$-dependence of the conduction-conduction and conduction-valence interlayer hops, which helps account for the differing effective masses in the subbands within the conduction band, is introduced as $t_{c(cv)}=t_{c(cv)}^{\Gamma}+t_{c(cv)}^{\prime}p^2$. Finally, $\Delta_{c(v)}$ are on-site energy shifts to the conduction(valence) states, included to take account of the different environment of states on the inside of the crystal compared with those on the surface.
\begin{table}
	\caption{HkpTB theory parameters in Eq. (\ref{full_ham}), and $\Gamma$ -point transition energies between two lowest subbands.\label{tab:ml_kp_parameters}}
	\begin{tabular}{cccc}
		\hline\hline
		$E_v$& $-2.79$~eV & $t^{\Gamma}_c$	& $0.34$~eV   \\ 
		
		$m_c$	&  $0.17~m_e$ & $t_v$	& $-0.41$~eV \\ 
		
		$\Delta_c$ & $0.03$~eV & $t_{cv}^{\Gamma}$	& $0.29$~eV  \\  
		
		$\Delta_v$ & $-0.03$~eV & $t_{c}^{\prime}$&  $-$5.91~eV\AA$^2$\\
		
		& & $t_{cv}^{\prime}$ & $-5.36$~eV\AA$^2$\\

		\hline 
	\end{tabular}
\quad
\begin{tabular}{cc}
	\hline\hline
	$N$ & $E_{2|N}-E_{1|N}$\\
	\hline
	2 & 680~meV \\
	3 & 490~meV \\
	4 & 360~meV \\
	5 & 280~meV \\
	\hline
\end{tabular}
\end{table}

We parametrize the interlayer hops ($t_{\alpha}$) and on-site energy shifts ($\Delta$) using dispersion curves obtained by means of density functional theory (DFT) as implemented in VASP\cite{VASP} for bulk and few-layer InSe\cite{magorrian2016electronic,soc2017}. The cutoff energy for the plane-wave basis is 600 eV and the Brillouin zone is sampled by a $12 \times 12$ $\mathbf{k}$-point grid. We complement DFT by a `scissor correction' adjustment of the monolayer band gap (having the effect $E_v\rightarrow E_v-0.99$~eV), chosen to correct for the difference between the LDA band gap and the value known from experiment for bulk InSe, as described in Ref. \onlinecite{magorrian2016electronic}. The parameters obtained are listed in Table \ref{tab:ml_kp_parameters}. This procedure is chosen since the underestimation of the gap by DFT would lead to the overestimation of the effect of the interband interlayer hop $t_{cv}$ on the value of the electron effective mass in the $z$-direction in the bulk, and on the subband spectra of multilayer films. To illustrate this effect, we consider the out-of-plane conduction band-edge effective mass in the bulk, given by Eq. (\ref{eq:mc_eff}). Using the parameters in Table \ref{tab:ml_kp_parameters} with the LDA band gap $E_g=0.41$~eV we obtain an effective mass $m_{Az}=0.043m_e$, while with the corrected gap $E_g=1.40~$eV we find an effective mass $m_{Az}=0.088m_e$, which is much closer to the experimental value of $0.081(9)~m_e$\cite{Kress-Rogers1982}. Having noted this change to the dispersion in the bulk crystal, we also expect changes to the energies of the subbands in the few-layer crystal. For example, correction of the monolayer gap reduces the splitting between the two lowest subbands in 6-layer InSe from 250~meV to 220~meV. 

\begin{figure*}
	\centering
	\includegraphics[width=1\linewidth]{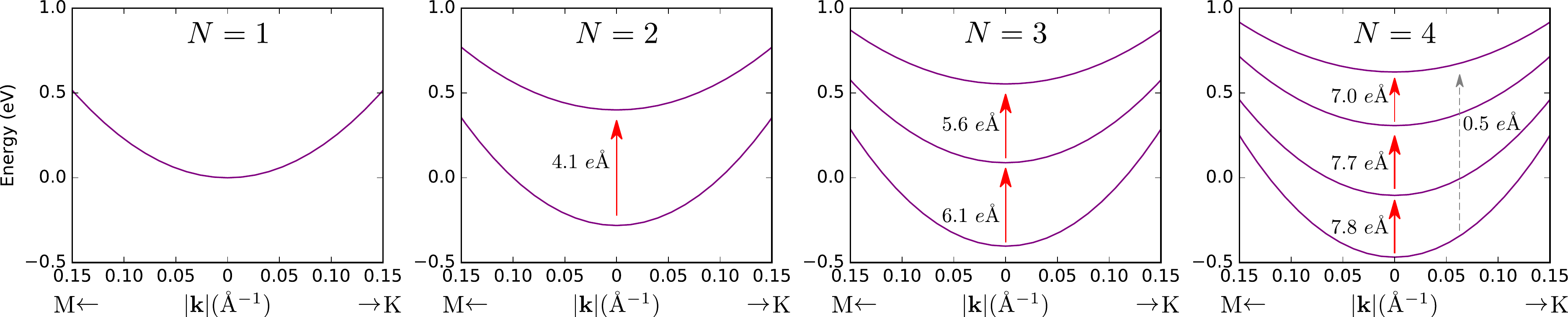}
	\caption{ Subbands of the conduction band in $N=1-4$-layer InSe near the $\Gamma$ point, from Eq. (\ref{full_ham}). 0 of energy set to conduction band minimum in the monolayer. Red arrows denote the strongest intersubband optical absorption transitions, coupled to the out-of-plane electric dipole, while the dashed gray arrow for 4-layer InSe indicates a much weaker transition. Arrows are labelled with the intersubband out-of-plane electric dipole moment of the transition, $d_z$ (Eq. (\ref{eq:dz})). }
	\label{fig:3lsb}
\end{figure*}

Each band in the monolayer generates $N$ subbands in $N$-layer InSe, with the subband dispersions of the conduction band for $N=1-4$ shown in Fig. \ref{fig:3lsb}, and the $\Gamma$-point separation between the lowest subbands shown in Table \ref{tab:ml_kp_parameters}. In all of these cases, electrons in the lower-energy subbands have lighter effective masses than those in the higher subbands. This difference in effective masses gives a finite thermal linewidth to the absorption lines, both at high doping and/or elevated temperatures.
\section{Band-edge expansion in bulk InSe}
In bulk InSe both conduction and valence band edges are located at the A-point (see Fig. \ref{fig:crys}), $k_x=k_y=0, k_z=\pi/a_z$ (where $a_z=8.32$~\AA~is the experimentally known interlayer distance\cite{Mudd2013}). The $\mathbf{k\cdot p}$ expansion in the vicinity of the A-point can be written as
\begin{equation}
\label{eq:bulk_kp}
E_c(\mathbf{p},p_z)=\left(\frac{\hbar^2}{2m_{A}}+\eta p_z^2a_z^2\right)p^2+\frac{\hbar^2p_z^2}{2m_{Az}}
\end{equation}
\noindent where $p=|\mathbf{p}|=|(p_x,p_y)|$, while $p_z=k_z-\pi/a_z$. The $xy$-plane and $z$-axis effective masses, $m_A$ and $m_{a_z}$, are given by
\begin{equation}
\label{eq:mc_eff}
\frac{1}{m_{A}}=\frac{1}{m_c}-\frac{4t_c^{\prime}}{\hbar^2},\quad \frac{1}{m_{A z}}=\frac{2a_z^2}{\hbar^2}\left[t^{\Gamma}_c+\frac{4t_{cv}^{\Gamma2}}{E_g}\right],
\end{equation}
respectively, where $E_g=2\Delta_c-(E_v+2\Delta_v)-2(t_c-t_v)$ is the bulk band gap. These give $m_A=0.11m_e$ and $m_{Az}=0.09m_e$, respectively, close to the experimentally known values of  $m_A=0.14m_e$ and $m_{Az}=0.08m_e$\cite{Kress-Rogers1982}. The additional parameter,
\begin{equation}
\eta=t_c'-\frac{2\hbar^2}{m_A}\frac{t_{cv}^{\Gamma 2}}{E_g^2}+\frac{8t_{cv}^{\Gamma}t_{cv}'}{E_g}\simeq -0.63\frac{\hbar^2}{2m_A},
\end{equation}
takes into account the anisotropic non-parabolicity of the electron dispersion at the A-point.

For a crystal slab of finite thickness $L=Na_z$ the general form of the boundary conditions at the crystal surfaces can be written as
\begin{equation}
\psi\pm \nu a_z\partial_z\psi=0,
\end{equation}
where $\nu$ is a dimensionless constant $\sim 1$, and allows the wavefunction to extend a little beyond the surface of the crystal. $+/-$ corresponds to the upper/lower surface of the crystal. Substitution of a general plane-wave wavefunction, $\psi=ue^{ip_zz}+ve^{-ip_zz}$, where $u$ and $v$ are constants, yields the requirement 
\begin{equation}
Np_za_z+2\arctan(\nu p_za_z)=n\pi,
\end{equation}
where $n$ is an integer. Expansion for small $p_z$ thus gives the quantization condition for small momenta,
\begin{equation}
p_z=\frac{n\pi}{(N+2\nu)a_z}.
\end{equation}
Within the bulk CB edge expansion, Eq. (\ref{eq:bulk_kp}), the 2D $\Gamma$-point energy of subband $n$ in N-layer InSe (denoted $n|N$) can then be expressed as
\begin{equation}
\label{eq:me_model_12}
E_{n|N}(n\ll N)\approx\frac{\hbar^2\pi^2}{2m_{A z}a_z^2}\frac{n^2}{(N+2\nu)^2}.
\end{equation}
Using subband energies calculated from the HkpTB model we find that $\nu=1.42$, as fitted to the inter-sub-band transition energies for the transition from subband 1 to 2, $E_{2|N}-E_{1|N}$. The energies obtained from Eq. (\ref{eq:me_model_12}) are plotted in Fig. \ref{fig:lw} alongside those obtained from the few-layer HkpTB model (Eq. (\ref{full_ham})). Additionally, the difference in effective masses for the electron dispersion in different subbands, shown in Fig. \ref{fig:3lsb}, arises from the non-parabolicity of the electron dispersion at the A-point. Also, quantization of $p_z$ in a thin film leads to heavier effective masses in higher subbands (for $n\ll N$),
\begin{equation}
\label{eq:me_ratio}
\frac{1}{{m_{n|N}}}\approx \frac{1}{m_A}\left[1-\frac{6.2n^2}{(N+2\nu)^2}\right],
\end{equation}
which produces the difference between the 2D effective masses in the lowest subbands shown in Fig. \ref{fig:lwf}.
\section{Intersubband transitions}
For the intersubband transitions between the subbands of the conduction band of n-doped InSe the population of holes in the valence band is negligible, so excitonic effects do not need to be considered, and the energy of an intersubband optical transition can be taken as that of the subband splitting. The oscillator strength of coupling to $z$-polarized photons is determined by the electric dipole matrix element,
\begin{equation}
\label{eq:dz}
d_z(1|N,b|N)=e\sum_n^N\bra{1|N}z(n)(a_{nc}^{\dagger}a_{nc}+a_{nv}^{\dagger}a_{nv})\ket{b|N},
\end{equation}
where $z(n)=a_z(n-(N+1)/2)$. Due to the $z\rightarrow -z$ symmetry of the HkpTB model $d_z(1|N,b|N)=0$ when $b$ is odd (a consequence of setting $t_{vc}=-t_{cv}$). Since the true crystal structure does not have this symmetry we check the validity of the latter assumption using values from a DFT calculation for the trilayer case - this gives $\frac{|d_z(1|3,3|3)|^2}{|d_z(1|3,2|3)|^2}\sim 10^{-4}$, so the transitions forbidden by the HkpTB model can be safely neglected. In Fig. \ref{fig:3lsb}, the non-zero intersubband dipole matrix elements are labelled alongside their respective transitions, and we note that the matrix element for transitions between adjacent subbands is much larger than that for transitions between more distant subbands. 

\begin{figure}
	\centering
	\includegraphics[width=0.95\linewidth]{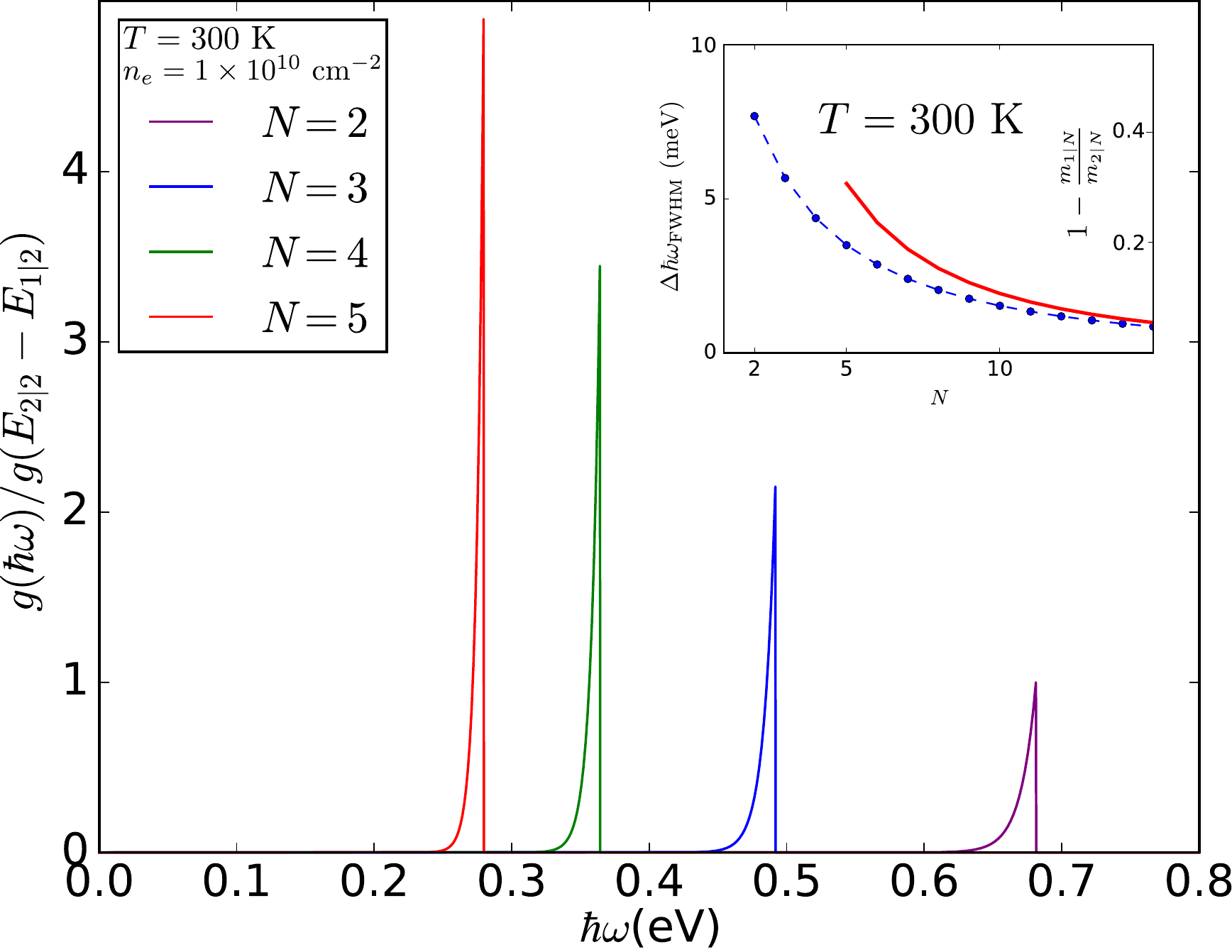}
	\caption{Intersubband lineshapes (normalized to the $\Gamma$-point transition in the bilayer) for $N$-layer InSe for excitation from lowest subband ($1|N$) to next-lowest subband ($2|N$), for a very light doping at $T=300$~K. Inset - thermal broadening ($T=300$K, left-hand axis) of absorption lines at light doping due to difference between subband effective masses (right-hand axis), Eq. (\ref{eq:linewidth}). Red line calculated using effective masses approximated by Eq. (\ref{eq:me_ratio}) for $N>4$.}
	\label{fig:lwf}
\end{figure}

With the subband energies, dipole matrix elements, and effective masses obtained by diagonalising the Hamiltonian in Eq. (\ref{full_ham}), we can describe the lineshape for intersubband absorption of IR/FIR photons by a slightly n-doped $N$-layer InSe, from the $n=1$ subband to the $n=2$ subband as
\begin{equation}
\label{eq:g}
g(\hbar\omega)\propto|d_z(1|N,2|N)|^2\hbar\omega\times\mathrm{DoS}\times F_T,
\end{equation}
where the joint density of states of the excitation is given by $\mathrm{DoS}(\hbar\omega)=\left[\pi\hbar^2\left(1/m_{1|N}-1/{m_{2|N}}\right)\right]^{-1}\times\Theta(E_{2|N}-E_{1|N}-\hbar\omega)$, while the factor reflecting the occupancy of initial states is
\begin{equation}
F_T=\left[\exp\left[\frac{1}{k_BT}\left(\frac{E_{2|N}-E_{1|N}-\hbar\omega}{1-\frac{m_{1|N}}{m_{2|N}}}-E_F\right)\right]+1\right]^{-1},
\end{equation}
where 
\begin{equation}
\label{eq:ef}
E_F=k_BT\ln\left[\exp{\left(\frac{\pi\hbar^2n_e}{m_{1|N}k_BT}\right)-1}\right]
\end{equation}
is the Fermi energy in the lowest subband, relative to the band minimum, of an n-doped InSe film with carrier density $n_e$. Here we assume that $E_{2|N}-E_{1|N}-E_F\gg k_BT$. The thermal linewidth can be estimated as
\begin{equation}
\label{eq:linewidth}
\Delta\hbar\omega_{\mathrm{FWHM}}\approx  \max\left\{\left[1-\frac{m_{1|N}}{m_{2|N}}\right]k_BT\ln{2},E_F\right\},
\end{equation}
resulting in the thermal linewidths shown in the inset to Fig. \ref{fig:lwf}, which shows the lineshapes (normalized to the $\Gamma$-point transition in the bilayer) determined by Eq. (\ref{eq:g}) for the $1|N\rightarrow 2|N$ IR/FIR optical transitions as a function of the transition energy for $N=2$ to $N=5$-layer InSe at 300K for a very light doping.
\section{Effects of interlayer screening in gated n-doped InSe}
In order for the intersubband transitions to be active the system must be n-doped. In the earlier transport experiments on 2D InSe, n-doping was introduced using electrostatic gates. In bulk systems (or thick films) doping by the gates induces accumulation layers of electrons near the surface, where the form of the confinement potential and, therefore, subband structure of the effective quantum well is determined by the density profile of confined electrons\cite{afs}. In a thin film, the doping by the gate applied on one side introduces an asymmetry of potential distribution inside it, increasing the energy separation between the lowest two subbands, while the change in the  corresponding lowest subband wave function leads to a partial screening of such potential. Below, we offer a self-consistent analysis of the potential profile and subband splittings induced by the voltage applied to the gate for doping the film with electrons, taking into account the screening (by the induced electrons) of electric field of  the gate. For this, we calculate the excess charges on each layer in the conduction band as
\begin{equation}
n_{e}(n)=\sum_j \frac{1}{\pi}\int \sum_{\alpha=c,v}|c_{jn}(\alpha,\hat{H'})|^2F_{Tj}(\hat{H'},k)k dk
\end{equation}
where $F_{Tj}(\hat{H'},k)$ are the Fermi occupation factors in the j-th subband at momentum ${\bf k}$, and $c_{jn}(c/v,H')$ are the amplitudes of the j-th subband wave function on the n-th layer (in terms of the monolayer basis states), evaluated using Eq. (\ref{full_ham}) with an additional potential energy term added to the on-layer `monolayer' Hamiltonian for each layer,
\begin{equation}
\hat{H}^{\prime}=\hat{H}+\sum_nU_n\left(a_{nc}^{\dagger}a_{nc}+a_{nv}^{\dagger}a_{nv}\right).
\end{equation}

The potential energy profile in $\hat{H}^{\prime}$ is related to the electron density distribution over the layers as
\begin{equation*}
U_{n>1}=U_1+ea_z\sum_{n^\prime=2}^{n}E_{n'-1,n'}, \quad E_{n-1,n}=\frac{e}{\varepsilon_0}\sum_{n^{\prime}=n}^{N}n_e(n^{\prime}),
\end{equation*}
which satisfies the requirement that the total density is determined by the electric field between the top of the film and the gate,
\begin{equation*}
E_{ext}=\frac{e}{\varepsilon_0}n_e,\quad n_e=\sum_n n_e(n). 
\end{equation*}
Then, for each density we converge the potential distribution $U_n$, setting an additional requirement that $U_1$ has a value chosen to give the desired total carrier density at self-consistency.
\begin{figure}
	\centering
	\includegraphics[width=0.99\linewidth]{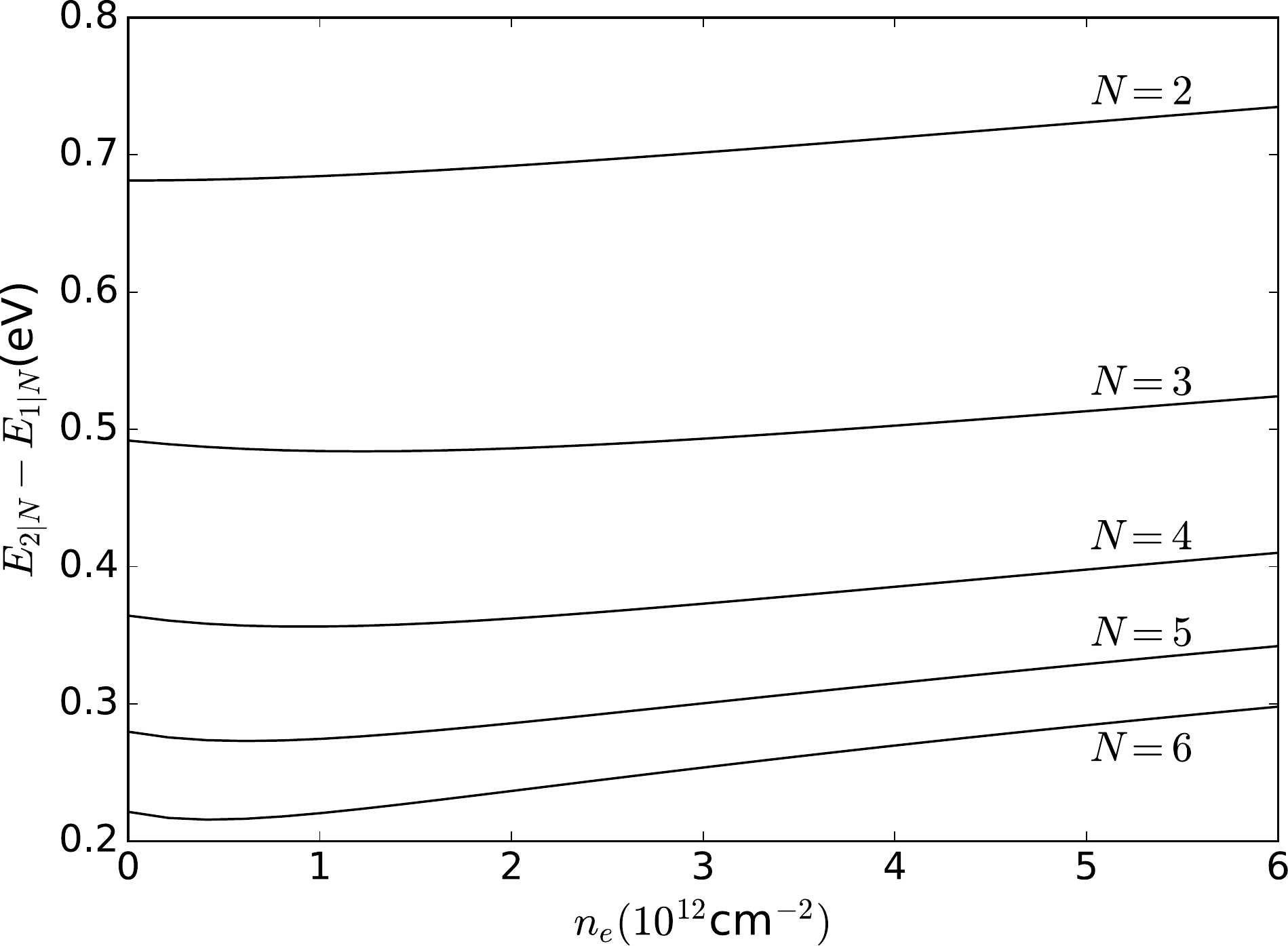}
	\caption{Intersubband transition energies as a function of total gate-induced carrier density ($n_e$) for 2-6 layer InSe.}
	\label{fig:screen}
\end{figure}

The results of the self-consistent calculation are shown in Fig. \ref{fig:screen} for the films with 2-6 layers, over the density range where only states in the lowest subband are filled. Following a slight decrease in the subband spacing at very small gate voltages (where the density distrubtion remains peaked in the center of the film) we find a steady increase in the intersubband transition energy. The latter result shows that by doping one can increase the intersubband spacing, thus broadening the spectrum of IR and FIR transitions in the film with a given number of layers, offering an additional tunability of the spectral characteristics of this system.

\section{Conclusions}
In conclusion, we have used a hybrid $\mathbf{k\cdot p}$-tight-binding model, fully parametrized using DFT, to evaluate the energies, oscillator strengths and thermal linewidths of optical transitions between the subbands of the conduction band of few-layer InSe. The strongest transitions are found to be from the lowest to next-lowest energy subbands, which broadly cover the the optical spectrum from $\sim$0.7~eV down to low THz range, with thermal linewidths $\sim 8 - 0.5$~meV at room temperature arising from the variation of in-plane effective masses between the subbands. Similar properties can also be expected for atomically thin films of transition metal chalcogenides\cite{TMD_subbands}, so that 2D materials offer great potential for applications in IR/FIR optoelectronics.
\begin{acknowledgements}
The authors thank S. Slizovskiy, A. Patan\`e, D. A. Bandurin, A. V. Tyurnina,  M. Potemski, Y. Ye, J. Lischner, and N. D. Drummond for discussions. This work made use of the CSF cluster of the University of Manchester. SJM and AC acknowledge support from EPSRC CDT Graphene NOWNANO EP/L01548X. VF acknowledges support from ERC Synergy Grant Hetero2D, EPSRC EP/N010345, and Lloyd Register Foundation Nanotechnology grant. VZ and VF acknowledge support from the European Graphene Flagship Project, the N8 Polaris service, the use of the ARCHER national UK supercomputer (RAP Project e547), and the Tianhe-2 Supercomputer at NUDT. Research data is available from the authors on request.
\end{acknowledgements}

%

\end{document}